\documentclass{appolb}
\usepackage{graphicx}
\usepackage{xcolor}
\usepackage{amsmath}

\usepackage{amsfonts} 
\begin{document}
% \eqsec  % uncomment this line to get equations numbered by (sec.num)
\title{Top-down approach to the curved spacetime effective field theory (cEFT) -- theory and examples
\thanks{Presented at the 6th Conference of the Polish Society on Relativity}%
% you can use '\\' to break lines
}
\author{{\L}ukasz Nakonieczny
\address{Institute of Theoretical Physics,
Faculty of Physics, University of Warsaw\\
ul. Pasteura 5, 02-093 Warszawa, Poland\\
e-mail: lukasz.nakonieczny@fuw.edu.pl}
}
\maketitle
\begin{abstract}
The effective field theory (EFT) turns out to be an instrument of an immense value in all aspects
of modern particle physics being theory, phenomenology or experiment. In the paper I will show
how to extend the systematic top down approach to construction of the EFT proposed by Hitoshi
Murayama (LBL, Berkeley) and separately by John Ellis (King’s Coll. London) groups to the curved
spacetime. To this end, I will take advantage of the heat kernel method so far extensively used in
obtaining the one-loop effective action in curved spacetime. After an introduction of the formalism
I will discuss its application to the problem of an influence of gravity on the stability of the Higgs
effective potential. 
\end{abstract}
\PACS{...}
  
\section{Introduction}
In the most rudimentary sense the Effective Field Theory is a tool that allows us to parametrize our 
lack of understanding. We use it when we do not have sufficient knowledge of details of investigated phenomena
or when the full description is too complicated to be tracked precisely, or when at the current state this 
precise description is not needed yet. The ways of construction of the EFT may be separated into two main 
categories, namely the bottom-up approach and the top-down approach. 
%In context of the high energy physics we use the bottom-up case when we are not interested in the detail of 
%the high energy part of the theory in other words we only want to have some parametrization of effects 
%induced by the high energy part that can be tested again experimental data. 
In the top-down approach we start with some extended theory that is presumably valid in the high energy region, 
then we integrate out the high energy degrees of freedom. In doing so we end up with the EFT 
in which effects associated with the presence of the integrated out particles are encoded in the 
higher dimensional operators. 

In the process of extending the EFT to the curved spacetime Effective Field Theory (cEFT) we followed 
the top-down approach. This contribution was based on \cite{Nak1}.

\section{Some technical aspects and results}

To extend the EFT to the cEFT, in other words to take into account the effects coming form the 
presence of the classical (non quantum) nontrivial spacetime curvature we decided to use 
the heat kernel method \cite{DeWitt}.
As an input to this method we need to have a classical action functional for the matter and gravity fields $S_{UV}$
that is valid at the high energy (UV stands for Ultra Violet). In our test case we used the system of two scalar fields
$H$ which may represent Higgs field and $X$ which represents heavy particles, for example heavy dark matter. 
The action is
\begin{align}
\label{SUV}
S_{UV} &= \int \sqrt{-g} d^4x \bigg \{
- \frac{1}{2} d_{\mu} H^{\dagger} d^{\mu} H - \frac{1}{2} m_{H}^2 |H|^2 
- \frac{\lambda_{H}}{4!} |H|^4 - \xi_H R |H|^2 + \nonumber \\
&- \frac{1}{2} d_{\mu} X d^{\mu} X - \frac{1}{2} m_{X}^2 X^2 - \xi_X R X^2 - \frac{1}{2} \lambda_{HX} X^2 |H|^2 
\bigg \},
\end{align}
where $d_{\mu}$ is a covariant derivative (it can also contain gauge field dependent part), $R$ is the Ricci scalar,
$m_{i}, \lambda_{i}$ and $\xi_{i}$, where $i=H,X$, are mass parameters, quartic couplings and non minimal couplings of the
scalar fields.    
In the next step our calculations follow closely the prescription of obtaining the one-loop effective action 
in curved spacetime \cite{BOS,Avr} with the added caveat that for now we consider only fluctuations of the heavy field $X$.
After this step we obtain the following action describing our low energy cEFT 
(for details please see \cite{Nak1}):
\begin{align}
\label{ScEFT}
S_{cEFT} &= \int \sqrt{-g} d^4x \bigg \{
- \frac{1}{2} d_{\mu} H^{\dagger} d^{\mu} H - \frac{1}{2} m_{H}^2 |H|^2 
- \frac{\lambda_{H}}{4!} |H|^4 - \xi_{H} R |H|^2 + \nonumber \\
&- \frac{1}{2} c_{dHdH} d_{\mu}|H|^2 d^{\mu}|H|^2 - c_{GHH} G^{\mu \nu} d_{\mu} |H|^2 d_{\nu}|H|^2 + \nonumber \\ 
&- c_{H} |H|^2  - c_{HH}|H|^4 - c_{6}|H|^6  \bigg \}.
\end{align}
The two last lines of (\ref{ScEFT}) represent contributions of the higher dimensional and gravity 
dependent operators to the physics of the Higgs field. 
   
In what follows we will focus on the gravity mediated contributions to the Higgs quartic coupling. 
They are described by the coefficient $c_{HH}$ in~(\ref{ScEFT}). These contributions are important form the standpoint of 
the vacuum stability of the Standard Model (SM). 
The $c_{HH}$ coefficient as calculated for the model described by the action (\ref{SUV}) is given by 
\begin{align}
c_{HH} &= \frac{\hbar}{(4 \pi)^2} \bigg [
\frac{\lambda_{HX}^2}{4 m_{X}^2} \left ( 2 \xi_X - \frac{1}{6} \right ) R 
- \frac{\lambda_{HX}^2}{8 m_X^4} \left ( 2 \xi_X - \frac{1}{6} \right )^2  R^2  + \nonumber \\ 
&- \frac{\lambda_{HX}^2}{720 m_{X}^4} \left ( \mathcal{K} - R_{\mu \nu} R^{\mu \nu} \right ) 
+ \frac{\lambda_{HX}^2}{m_{X}^4} \left ( - \frac{1}{4} \xi_X + \frac{1}{40} \right ) \square R 
- \frac{\lambda_{HX}^2}{90 m_{X}^4}  \nabla_{\mu} \nabla_{\nu} R^{\mu \nu} 
\bigg ],
\end{align} 
where $\mathcal{K} \equiv R_{\mu \nu \rho \sigma} R^{\mu \nu \rho \sigma}$ is a Kretschmann scalar and 
$\square \equiv \nabla_{\mu} \nabla^{\mu}$ is a covariant d'Alembert operator. 

\begin{figure}[htb]
\centerline{%
\includegraphics[width=12.5cm]{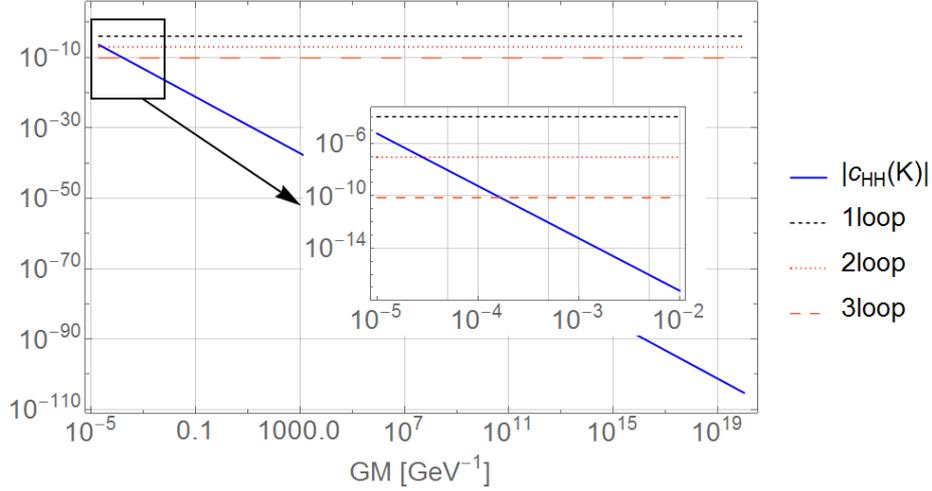}}
\caption{Gravity induced contribution to the Higgs quartic coupling in the black hole background. 
$|c_{HH}(K)| = |- \frac{1}{(4 \pi)^2} \frac{\lambda_{HX}^2}{720} \frac{\mathcal{K}}{m_{X}^4}|$,
$G$ is the Newton constant, $M$ is the black hole mass and loops prefactors are given by the formula 
nloop$= \frac{\lambda_{H}^{n+1}}{(16 \pi^2)^n}$. For the plot we chose $\lambda_{HX} = 0.25$, $\lambda_{H} = 0.13$
and $m_{X} = 10 {\rm TeV}$.}
\label{Fig:1}
\end{figure}

\begin{figure}[htb]
\centerline{%
\includegraphics[width=12.5cm]{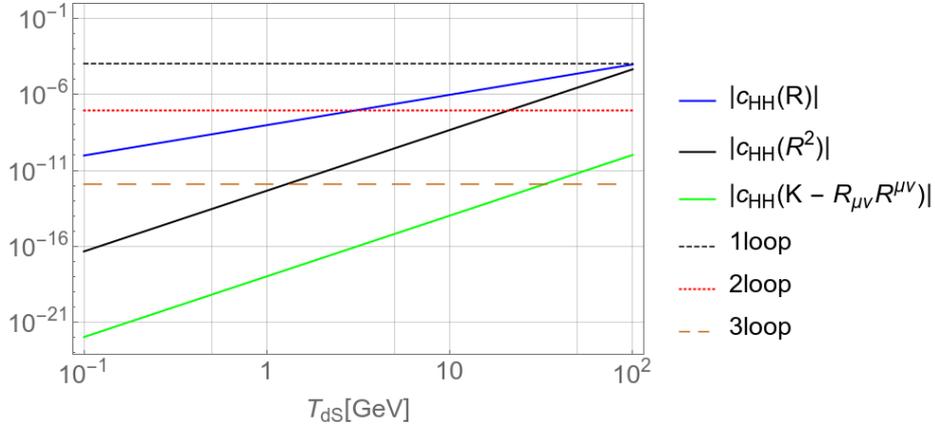}}
\caption{
Gravity induced contribution to the Higgs quartic coupling in the de Sitter like FLRW background. 
Loops prefactors are given by the formula nloop$= \frac{\lambda_{H}^{n+1}}{(16 \pi^2)^n}$ and 
$T_{dS}$ is the temperature of the de Sitter spacetime. 
For the plot we chose $\lambda_{HX} = 0.25$, $\lambda_{H} = 0.13$, $m_{X} = 10 {\rm TeV}$ and $\xi_X = 10$.
For comparison $T_{\odot} \sim 10^{-13} \textrm{GeV}$, and $T_{EW} \sim 10^2 \textrm{GeV}$.}
\label{Fig:2}
\end{figure}

The magnitude of this contribution to the Higgs quartic coupling for two selected 
fixed gravitational backgrounds are depicted in Figs.~\ref{Fig:1} and \ref{Fig:2}.

%uncomment the following lines to place a figure

The first case is the one of the primordial black hole (PBH) modeled by the Schwarzschild metric. 
From Fig.~\ref{Fig:1} we see that for small black hole mass $M_{PBH} \sim 10^{10} \textrm{g}$
the gravity mediated contributions are comparable to the two-loop effects coming form the Higgs 
self-interaction. This implies that they should be taken into account if we want to 
analyze the problem of the SM vacuum stability near such a black hole beyond the one-loop approximation.

The second case is the gravity mediated contribution in the de Sitter spacetime.
Physically this type of spacetime may describe short post inflationary reheating stage of the evolution 
of the Universe. From Fig.~\ref{Fig:2} we may see that if the temperature at this epoch 
is of the order of electroweak phase transition ($T_{EW}$) then 
gravity mediated effects are of the order of the two-, or even close to the one-loop effects
and they should be included in the analysis of the vacuum stability.

\section*{Acknowledgements}
{\L}N  was supported by the National Science Centre, Poland under a grant DEC-2017/26/D/ST2/00193.

\end{document}